\documentclass{iau} 
\usepackage{graphicx}

\title[Pulsar Science with LEAP] 
{Pulsar science with data from the Large European Array for Pulsars}

\author[James W. McKee]   
{James W. McKee$^{1}$ on behalf of the LEAP group$^{\dagger}$}

\affiliation{$^1$Max-Planck-Institut f{\"u}r Radioastronomie, Auf dem H{\"u}gel 69, D-53121 Bonn, Germany \\ email: {\tt jmckee@mpifr-bonn.mpg.de}} 

\pubyear{2017}
\volume{337}  
\jname{Pulsar Astrophysics -­ The Next 50 Years}
\editors{P. Weltevrede, B.B.P. Perera, L. Levin Preston \& S. Sanidas, eds.}

\begin{document}

\maketitle

\begin{abstract}
{\let\thefootnote\relax\footnote{{$^{\dagger}$C.\,G.\,Bassa, S.\,Chen, M.\,Gaikwad, G.\,H.\,Janssen, R.\,Karuppusamy, M.\,Kramer, K.\,J.\,Lee, K.\,Liu, D.\,Perrodin, S.\,A.\,Sanidas, R.\,Smits, B.\,W.\,Stappers, L.\,Wang, and W.\,W.\,Zhu}}}
The Large European Array for Pulsars (LEAP) is a European Pulsar Timing Array project that combines the Lovell, Effelsberg, Nan{\c c}ay, Sardinia, and Westerbork radio
telescopes into a single tied-array, and makes monthly observations of a set of millisecond pulsars
(MSPs). The overview of our experiment is presented in \cite[Bassa \etal\ (2016)]{Bassa_etal16}. Baseband data are recorded at a central frequency of 1396\,MHz and a bandwidth of 128\,MHz
at each telescope, and are correlated offline on a cluster at Jodrell Bank Observatory using a purpose-built correlator, detailed in \cite[Smits \etal\ (2017)]{Smits_etal17}. 
LEAP offers a substantial increase in sensitivity over that of the individual telescopes, and can operate in timing and imaging modes (notably in observations of the galactic centre radio magnetar; \cite[Wucknitz 2015]{Wucknitz15}). 
To date, 4 years of observations have been reduced. Here, we report on the scientific
projects which have made use of LEAP data.
\end{abstract}

\firstsection 
\section{PSR\,B1937+21 Giant Pulses}
\begin{figure}
\begin{center}
 \includegraphics[scale=0.115]{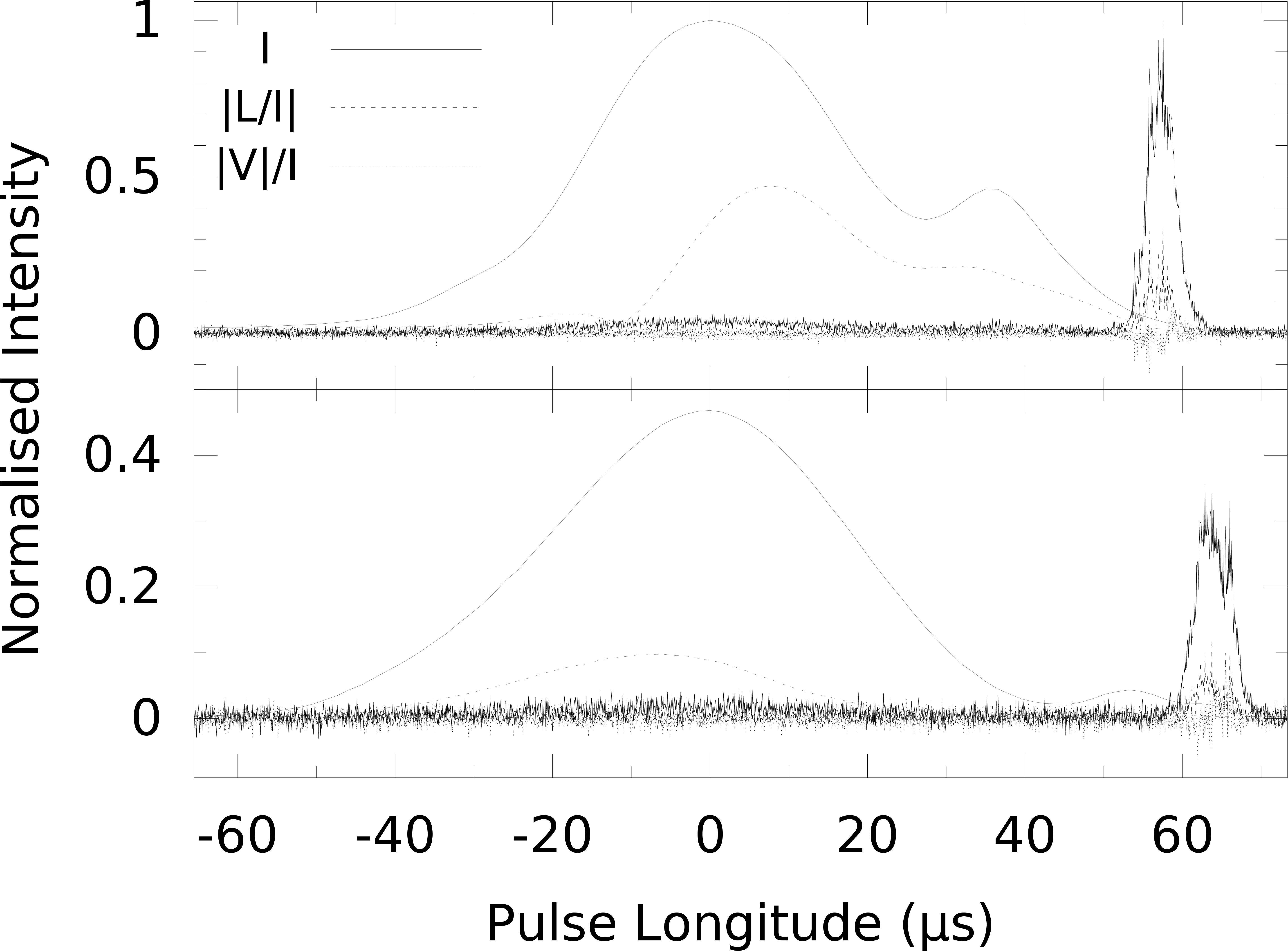} 
 \includegraphics[scale=0.707]{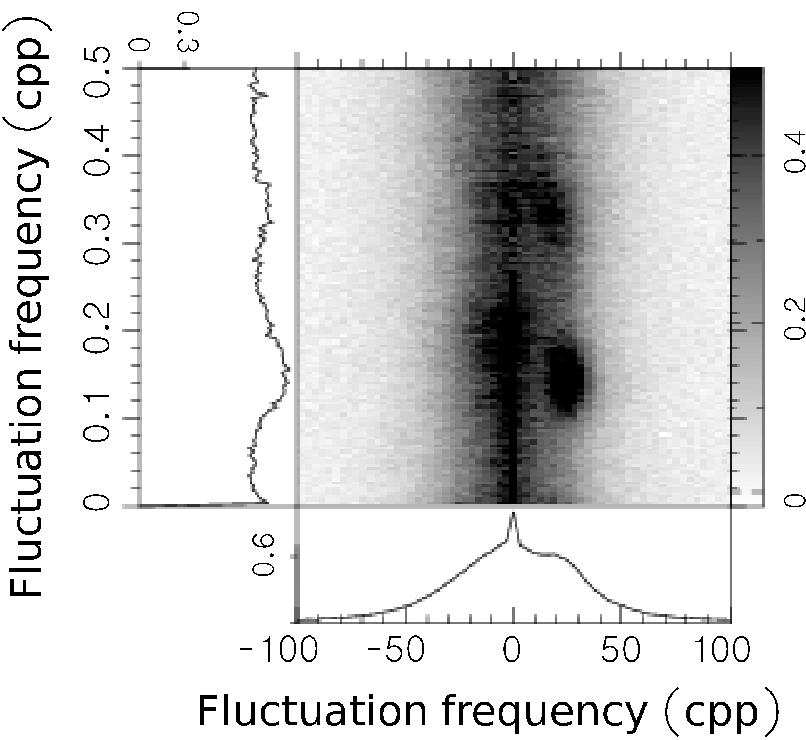} 
 \caption{\textit{left:}\,PSR\,B1937+21 Stokes parameters vs. phase of the regular emission (grey) and GP emission (black), averaged over all our observations. The GP emission occurs at the edge of the regular emission regions,
with a lag of 58\,$\mu$s relative to the main pulse and 64\,$\mu$s to the interpulse, and in
both plots we see a small contribution from the regular emission is present in
the GP data, indicating that regular emission occurs simultaneously with GP
emission.
\textit{right:}\,PSR\,J1713+0747 2D fluctuation spectrum, with horizontal and vertical integrations. 
Two maxima occur at 0.14 and 0.34\,cpp, confirming phase-dependent modulation.}
   \label{fig1}
\end{center}
\end{figure}

Giant pulses (GPs) are a rare phenomenon (see in only 11 pulsars) where occasional single pulses
of flux densities orders of magnitude greater than the mean are emitted.
We have searched for GPs in $3.1\times10^{7}$ rotations of PSR\,B1937+21 with LEAP,
detecting 4265 GPs (Figure \ref{fig1}); the largest ever sample of GPs gathered for
this pulsar. We have examined the distribution of polarisation fractions
of GPs, finding no correlation between GP flux and fractional
polarisation, and no correlation between polarisation fraction and pulse
phase. We have measured the power law index of the flux distribution to
be $\alpha = -3.8\pm0.2$, and have noted a low-flux turnover at $\sim4$\,Jy. We have
measured modulation indices for the GPs, and find that they vary by
$\sim0.5$ towards the centre of the phase distributions, in contrast to the
findings of \cite[Jenet \etal\ (2001)]{Jenet_etal01} for the highly-stable regular emission. We compare the
timing prospects of PSR\,B1937+21 using GPs and the average profile
separately, and find no improvement when using the GPs for timing. This work is included in \cite[McKee (2017)]{McKee17}, with a paper in preparation.

\section{M28 Pulsar Search}
Globular clusters (GCs) are dense regions of stars, tightly
bound to a gravitational centre. The high stellar density of
GCs makes binary formation more likely, giving rise to large
populations of binary pulsars and MSPs, and increases the
probability of exotic binary systems forming (e.g. pulsar-black hole binaries). For this reason, GCs are attractive
targets for pulsar searches (e.g. \cite[Hessels \etal\ 2007]{Hessels_etal07}).
The tied-array beam of LEAP is small (max. size: 50\,mas),
and therefore impractical for blind pulsar surveys, but
the sensitivity of LEAP makes it well-suited to targeted
searches. We are searching for pulsars in GC\,M28, by first phasing up on the bright MSP B1821$-$24A (M28A), and then
adjusting the measured phase delays to reposition the virtual LEAP beam. As
the angular size of M28 (1.56\,arcmin) is much larger
than our tied-array beam, we focus our search on currently-unclassified
X-ray sources within the cluster, which may be associated with unknown MSPs. The ongoing search has produced re-detections of known pulsars within the GC, and will be the subject of an upcoming paper (Wang \etal\, in prep.).

\section{PSR\,J1713+0747 Single Pulses}
Single-pulse studies offer an opportunity to understand rotational variability of pulsars. These studies
require bright pulsars and highly sensitive instruments,
in order to discriminate single pulses from the
background noise.
We have performed a single-pulse study of PSR\,J1713+0747, using $197,000$ pulses
from when the pulsar was brightly scintillating.
Using these data, we confirm the detection of periodic
intensity modulation, discovered by \cite[Edwards \& Stappers (2003)]{Edwards_etal03}, which we demonstrate to be phase-dependent (Figure \ref{fig1}); the first detection of such behaviour in an
MSP. The drifting sub-pulses are found to have two
modes, with $P_{2} = 6.9\pm0.1P$, and $P_{3} = 2.9\pm0.1P$, where $P$ is the spin period. We find
that the fractional polarisation scales with flux density,
with the brightest pulses being highly linearly-polarised. This work has been published in \cite[Liu \etal\ (2016)]{Liu_etal16}.

\end{document}